\documentclass[aps,prb,amsmath,amssymb,reprint,superscriptaddress,longbibliography]{revtex4-2}
\usepackage{graphicx}
\usepackage{dcolumn}
\usepackage{wasysym}
\usepackage{bm}
\usepackage{color}
\usepackage{float}
\usepackage[utf8]{inputenc}
\usepackage[T1]{fontenc}
\usepackage{etoolbox}

\begin{document}


\title{\textit{In-vivo} blood pressure sensing with bi-filler nanocomposite}

\author{Chandrabhan Kushwah}
\affiliation{Department of Physics, University of Bath, Bath BA2 7AY, United Kingdom}

\author{Martin Riesenhuber}
\affiliation{Department of Medicine, University of Vienna, Spitalgasse 23, 1090 Vienna, Austria}

\author{S\o ren Asmul}
\affiliation{Medtronic Bakken Research Centre, Endepolsdomein 5, The Netherlands}

\author{Mariann Gy\"ongy\"osi}
\affiliation{Department of Medicine, University of Vienna, Spitalgasse 23, 1090 Vienna, Austria}

\author{Alain Nogaret}
\email{A.R.Nogaret@bath.ac.uk}
\affiliation{Department of Physics, University of Bath, Bath BA2 7AY, United Kingdom}

\begin{abstract}
Conductive elastomers present desirable qualities for sensing pressure \textit{in-vivo}, such as high piezoresistance in tiny volumes, conformability and, biocompatibility.  Many PDMS-based electrically conductive nanocomposites however, are susceptible to electrical drift following repeated stress cycles and chemical aging.  Innovative approaches are needed to stabilize their percolation network against deformation to improve reproducibility and facilitate sensor calibration.  One approach we propose here is to decouple the tunnelling-percolation network of HOPG nanoparticles from the incomplete viscoelastic recovery of the PDMS matrix by inserting minute amounts of insulating SiO$_2$ nanospheres.  SiO$_2$ nanospheres effectively reduce the number of nearest neighbours at each percolation node switching off the parallel electrical pathways that might become activated by incomplete viscoelastic relaxation.  We varied the size of SiO$_2$ nanospheres and their filling fraction to demonstrate nearly complete piezoresistance recovery when SiO$_2$ and HOPG nanoparticles have equal diameters ($\approx$400nm) and SiO$_2$ and HOPG volume fractions are 1\% and 29.5\% respectively.  We demonstrate an \textit{in-vivo} blood pressure sensor based on this bi-filler composite.
\\
\\
\noindent \textit{Highlights}:
\begin{itemize}
  \item Carotid arterial pressure is measured with a bifiller nanocomposite
  \item Added silica nanobeads improve sensitivity to pressure
  \item Silica nanobeads stabilise sensor against incomplete viscoelastic recovery
  \item Sensor is scalable, biocompatible and has low fabrication cost
\end{itemize}
\end{abstract}

\maketitle

\section{Introduction}

Bioelectronic medicine is driving the search for minimally invasive sensors capable of feeding back physiological signals to biomedical implants.  Pressure sensing in particular is needed for the treatment of chronic diseases that rely on monitoring blood pressure~\cite{Trohman2020,Ruth2021} and lung inflation~\cite{Shanks2022,Nogaret2015,Callaghan2020}.  Existing solid-state pressure sensors exploit optical~\cite{Fiala2012}, triboelectric~\cite{Ma2016}, piezoresistive~\cite{Peng2021,Barlian2009,Lee2016b}, capacitive~\cite{Kim2019,Dehe1995}, and electromechanical~\cite{Lee2013,Inoue2019,Potkay2008} transduction.  These sensors tend to be used acutely inside catheters~\cite{Cleven2014} but are less well suited to chronic use.  Conductive elastomers, on the other hand, offer several advantages including high piezoresistive gauge factors, conformability and, biocompatibility which makes them good candidates for implants~\cite{Zang2015,Boland2014,Gaumet2021,Meier2007,Littlejohn2013, Negri2010,Someya2004,Chen2007}.  PDMS nanocomposites filled with conductive nanoparticles can in principle detect the pulsatility of blood vessels and changes in blood pressure through changes in the resistance of their percolation network.  The nanoparticles are globally connected to one another by exponentially decaying tunnelling probabilities~\cite{Balberg2009}.  Their conductivity follows a scaling law $\sigma \sim (\nu-\nu_c)^\alpha$~\cite{Grimaldi2006,Balberg2009,Feng1987,Johner2007,Johner2008} with a conductance threshold $\mu_c$~\cite{Bruggeman1935,Landauer1952,Vionnet2005} equal to $\nu_c=24\%$ for 450nm HOPG nanoparticles in PDMS~\cite{Littlejohn2011,Chauhan2017}.  Near the percolation threshold, the piezoresistance gauge factor $GF \propto \nu/(\nu-\nu_c)$ can have a large value~\cite{Taylor2017}.  The response of PDMS nanocomposites to dynamical loading was modelled under both hydrostatic pressure~\cite{Grest1986} and uniaxial strain~\cite{Taylor2017}.  The viscoelastic response of the PDMS matrix~\cite{Hunter1976} was found to introduce relaxation times in the electrical conductivity such as the stress relaxation time ($\tau_R$), the creep time ($\tau_c$) and the possibility of incomplete recovery ($\epsilon$) that accounts for the built up of internal strain~\cite{Taylor2017}.  The latter introduces irreversible changes in the topology of the percolation network which causes the resistance drift that we seek to minimize.  In order to improve reversibility, several groups have attempted to microengineer the percolation network by controlling the orientation of carbon nanotubes~\cite{Zhang2007,Pang2012,Lipomi2011} or nanoplatelets~\cite{Mannsfeld2010,Nasirpouri2015}.  Recently Ibrahim et al.~\cite{Ibrahim2012} have sought to control the mechanical properties of PDMS by inserting insulating nanospheres.

\begin{figure*}[ht]
\includegraphics[width=0.8\linewidth]{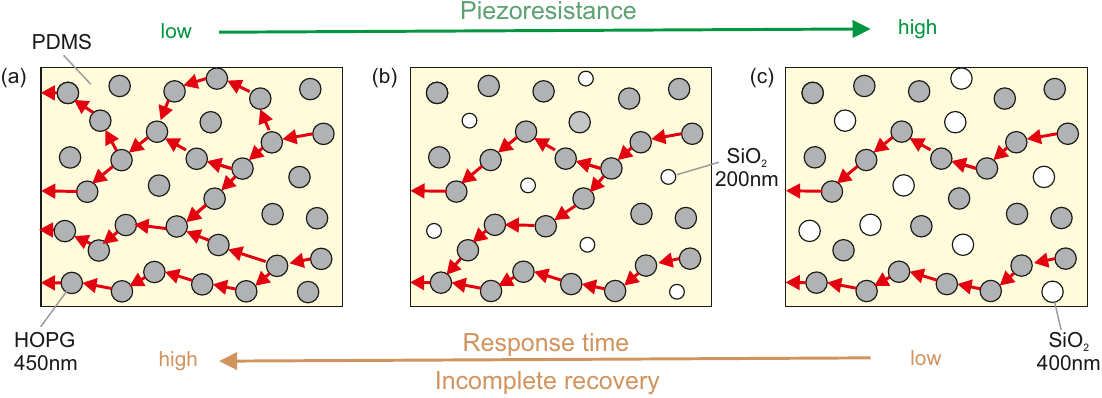}\\
\caption{\textbf{Effect of SiO$_2$ nanoparticles on the tunnelling-percolation network}
\\
(a) PDMS matrix filled with 450nm HOPG nanoparticles (grey dots).  Percolation pathways (red arrows) branch out at each node.  The different branches compete to carry the current, giving a high conductivity, fast piezoresistance response, but incomplete viscoelastic recovery.  (b) Bi-filler composite mixing 200nm SiO$_2$ nanoparticles with 450nm HOPG in PDMS.  (c)  Bi-filler composite mixing 400nm SiO$_2$ nanoparticles with 450nm HOPG in PDMS.  The SiO$_2$ nanospheres reduce the number of electrical pathways crossings at each HOPG node effectively decoupling the electrical response of the percolation network from viscoelastic relaxation.}
\label{fig:fig1}
\end{figure*}

In this paper, we report on the piezoresistance of bi-filler composite ribbons and evaluate their performance as implantable blood pressure sensors.  We studied the effect of dispersing silica nanospheres in various concentrations and diameters in a conductive elastomer embedding HOPG nanoparticles in PDMS.  The volume concentration of HOPG nanoparticles was $\nu=29.5\%$ and their diameter was $d_c=450$nm in all samples while we varied the diameter of SiO$_2$ nanospheres ($d_i=$200nm, 400nm) and their concentrations (0\%, 1\% and 2\%). SiO$_2$ nanospheres were found to improve the piezoresistance recovery with the figure of merit for incomplete recovery dropping from $\epsilon$=38.2\% without SiO$_2$, to $\epsilon$=1.52\% with 1\% SiO$_2$, and $\epsilon=1.20\%$ with 2\% SiO$_2$.  The piezoresistance also increased with increasing SiO$_2$ filling fraction.  Inserting larger SiO$_2$ nanospheres increased the response time of the resistance to pressure, from 0.2s without SiO$_2$ to 0.28s for $d_i=200$nm, and 0.30s for $d_i=400$nm.  We then designed a blood pressure sensor that converted arterial pressure into a piezoresistance change.  The piezoresistance change was calibrated with a pressure gauge.  \textit{In-vivo} trials show that the bi-filler composite sensor successfully reads beat-to-beat pressure oscillations on the carotid artery of pigs.  In each cardiac cycle, the sensor resolved systole and diastole pressure patterns with excellent accuracy.  The sensor also picked up the baseline increase in blood pressure that followed the injecting of a vasoconstrictor.  Our sensor sensitivity is at least equal to that of external pressure sensor.  Bi-filler composites thus provide a superior sensitivity for in-vivo pressure monitoring with a low resistance drift.

\section{Materials and Methods}

\subsection{Bifiller composites}

We mixed HOPG and SiO$_2$ nanoparticles in PDMS with two working hypotheses.  The first was that SiO$_2$ nanospheres provide a way to control the Young's modulus $E$~\cite{Ibrahim2012} of the composite independently of its electrical properties.  Simultaneously, the nanospheres would reduce the effective HOPG filling fraction, bringing $\nu$ closer to the percolation threshold, $\nu_c$, hence increasing the piezoresistance.  The second hypothesis was that SiO$_2$ nanoparticles would reduce the number of closest neighbours of HOPG nanoparticle.  This effectively prunes out the most tortuous percolation paths which are the most sensitive to irreversible alterations in the topology of the network upon strain cycling.  This would thus improve reproducibility of resistance upon stress cycling.

Figure~\ref{fig:fig1} schematically depicts the effect of SiO$_2$ nanoparticles in stabilizing percolation paths.  In composites incorporating HOPG nanoparticles only, percolation paths branch out in all directions forming parallel channels that compete to carry the current (Fig.\ref{fig:fig1}a).  The insertion of SiO$_2$ nanoparticles 200nm (Fig.\ref{fig:fig1}b) and 400nm in diameter (Fig.\ref{fig:fig1}c) selectively switches off the longer percolation paths.  These are no longer able to compete with the low resistance paths when the topology of the network does not return to the exact same initial state after a stress cycle.

\subsection{Composite preparation}

\begin{table}[ht]
\begin{center}
\begin{tabular}{|c|c|c|c|c|c|}
\hline \hline
& \multicolumn{5}{c|}{Bifiller composite} \\
& \multicolumn{5}{c|}{PDMS +} \\
& \multicolumn{5}{c|}{29.5\% HOPG 450nm +}  \\
& \multicolumn{2}{c|}{SiO$_{2}$ 400nm} & \multicolumn{2}{c|}{SiO$_{2}$ 200nm} & - \\
& \;\; 1\% \;\; & \;\; 2\% \;\; & \;\; 1\% \;\; & \;\; 2\% \;\; & \; \; - \; \; \\
\hline
Catalyst mass (g) & 0.1 & 0.1 & 0.1 & 0.1 & 0.1 \\
PDMS mass (g) & 2 & 2 & 2 & 2 & 2 \\
HOPG mass (g) & 1.733 & 1.733 & 1.733 & 1.733 & 1.733 \\
SiO$_{2}$ mass (g) & 0.060 & 0.121 & 0.066 & 0.134 & 0 \\
\hline \hline
\end{tabular}
\end{center}
\caption{\textbf{Compositions of bi-filler composites}  \\  The \% indicate concentrations in volume of SiO$_2$ and HOPG in liquid silicone (Alchemie RTV137).  The bifiller composite mixture was cured at 300K with the catalyst (Alchemie C137S).}
\label{tab:tab1}
\end{table}

Our reference PDMS nanocomposite was filled with a 29.5\% volume fraction of HOPG nanoparticles, $d_c=450$nm in diameter.  This was sufficiently close to the 24\% percolation threshold to give a high piezoresistance that was consistent among all fabricated samples.  The average tunnelling barrier width, $b$, and the average number of nearest neighbours, $B$, of each HOPG nanoparticle was first estimated before mixing in SiO$_2$ nanospheres.  We equated the volume density of HOPG nanoparticles $N=29.5\%\;3/(4\pi)(d_c/2)^{-3}$ to the volume density of randomly distributed closed-packed spheres $N=64\%\;3/(4\pi)r_{cp}^{-3}$ of radius $r_{cp} = (d_c + b)$.  Using 64\% as the volume fraction of closed packed spheres, we find $r_{cp}=1.295 (d_c/2)$ and $b=2r_{cp}-d_c=0.589(d_c/2)$.  For a HOPG diameter $d_c=450$nm, the average tunnelling barrier is $b=265$nm.

A similar calculation done at the percolation threshold ($\nu_c=24\%$) yields a critical tunnelling barrier of 348nm~\cite{Balberg2009}.  From this, we obtain the mean critical radius $r_c=1.387a$ which determines the topologically required number of nearest neighbours, $B_c$, to achieve a globally connected network.  Ambegaokar et al~\cite{Ambegaokar1971} have showed that this is given by $B_c=(32\pi/3)(r_c^3-(d_c/2)^3)N$ from which we obtain $B_c=3.2$.  This value is good agreement with the 2.8 nearest neighbours theoretically predicted at the onset of conduction for a 3D composite filled with metallic nanoparticles~\cite{Balberg2009}, thus validating our mean tunnelling barrier width estimates.  At the 29.5\% HOPG volume fraction, the average number of nearest neighbours is evidently greater that 3.2, however this value will drop back towards 3.2 once SiO$_2$ nanospheres are mixing in.

We have prepared the five types of composites listed in Table~\ref{tab:tab1}.  The nanoparticles were vigorously mixed in liquid RTV silicone using pestle and mortar for several minutes past the point where the mixture had become homogeneous.  The catalyst was then added and the mixture was cast in a ribbon-shaped extrusion mould (Fig.\ref{fig:fig2}a) laid on a pristine (non-conductive) PDMS film.  The composite was left to cure at room temperature for over 48 hours.  The piezoresistive ribbon was 3mm wide and 0.7mm thick.  Au contacts (500nm thick) were thermally evaporated over the ribbon pads leaving a 10mm exposed section of composite at the centre of the ribbon which curled around the blood vessel.  The gold pads had a contact resistance of ~1$\Omega$~\cite{Littlejohn2011} which was negligible compared the resistance of 10mm active central section (2-60k$\Omega$).  The device was finally capped with a thin pristine PDMS layer leaving only via-holes over contact pads (grey areas, Fig.\ref{fig:fig2}b).  The top and bottom pristine PDMS layers encapsulated the nanocomposite ribbon inside an electrically insulating, biocompatible sheath.

\subsection{Device design}

\begin{figure}
\includegraphics[width=\linewidth]{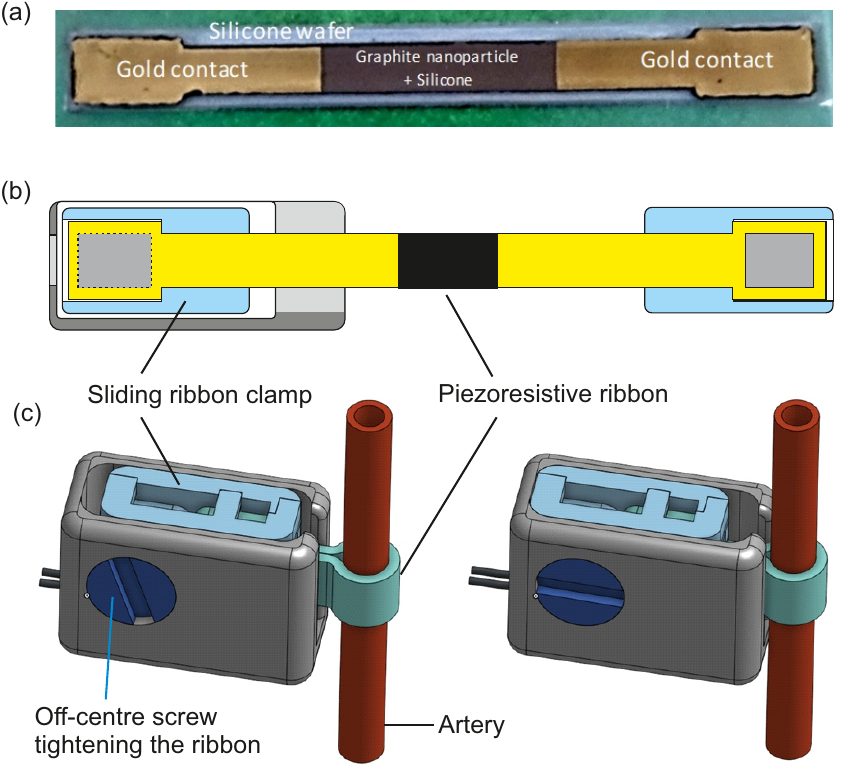}\\
\caption{\textbf{Blood pressure sensor}
\\
(a) Piezoresistive ribbon (3mm wide, 0.7mm thick).  The composite (black layer) stands on a pristine PDMS substrate (white underlayer) and is capped at its extremities by Au contact pads.  A transparent pristine PDMS overlayer provides insulation and biocompatibility.  (b) Schematics of the piezoresistive ribbon held by sliding clamps (blue elements) that fold onto each other and fit in the sensor casing (grey box).  Via holes through the PDMS overlayer (grey rectangle) allow connecting wires to Au pads.  (c) Sensor mount around the blood vessel.}
\label{fig:fig2}
\end{figure}

The active area of the ribbon was wrapped around the blood vessel (Fig.\ref{fig:fig2}c) and was held in place by two 3D printed elements that clamped the ends of the ribbon and prevented it from slipping (Fig.\ref{fig:fig2}c).  The clamps were then joined together and inserted in the 3D printed housing.  The clamps were allowed to slide by up to 2mm in the housing to tighten the ribbon around the blood vessel.  The clamp position was adjusted with an off-centre peg which was inserted through a lateral hole in the housing and plugged into a hole in the clamps.  The screw peg was conically shaped, like a violin peg, so that the ribbon tension could be adjusted with less than half a turn and held in position by simply pushing the peg in the hole.  The clamping assembly was sterilised before implanting.

Electrical wires were connected to the ribbon endings using silver epoxy.  The ribbon resistance was measured in a two-point configuration under a 1$\mu$A constant current.  The voltage across the ribbon was read by a 16 bit analogue-to-digital converter and wirelessly transmitted to the monitoring laptop with a bluetooth protocol.

\subsection{Calibration bench}

\begin{figure}[ht]
\includegraphics[width=\linewidth]{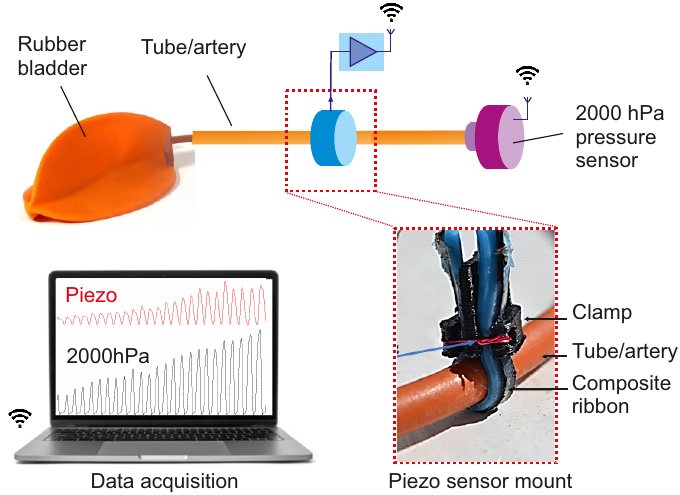} \\
\caption{\textbf{Piezoresistance measurements}
\\
Piezoresistive ribbon wrapped around a rubber tube simulating the carotid artery.  A rubber bladder was used to vary the inside pressure.  The other end of the rubber tube had a 2000hPa pressure gauge to calibrate the piezoresistive sensor.  Piezoresistance and pressure signals were acquired in real time by a CASSY interface (laptop).}
\label{fig:fig3}
\end{figure}

The response of the five elastomers in Table~\ref{tab:tab1} was investigated in the lab to identify the bi-filler composition best suited for trials.  We built the experimental bench in Fig.\ref{fig:fig3} to simulate the sensor response to changes of pressure in an artery.  The artery was modelled by a section of rubber tubing with mechanical properties and dimensions similar to the carotid artery.  A rubber bladder closed the tube at one end.  A 2000hPa pressure sensor closed the tube at the other end for the purpose of calibrating the piezoresistance.  The rubber bladder was compressed and released with increasing strength producing 25 resistance peaks of increasing magnitude followed by 25 peaks of decreasing magnitude.  Both pressure and resistance were recorded simultaneously and plotted on the laptop screen with the CASSY-2 acquisition software (Fig.\ref{fig:fig3}).

The piezoresistance was derived from the measurements of the resistance peak amplitudes as a function of pressures.  The sensor response time was obtained from the time delay between the piezoresistance peak and the peak pressure averaged over the 50 oscillations.  The incomplete recovery ratio was calculated as $\epsilon=|R_{end}-R_{start}|/R_{start}$ when $R_{start}$ and $R_{end}$ are the resistance readings at the start and end of the series.

\subsection{\textit{In-vivo} measurements}

\begin{figure}[ht]
\includegraphics[width=0.8\linewidth]{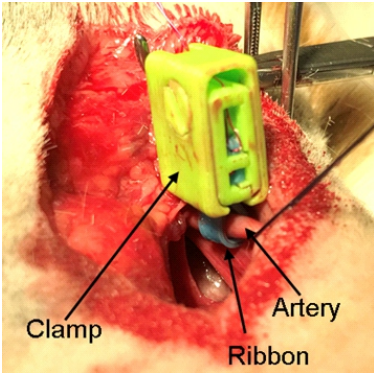}\\
\caption{\textbf{In-vivo implant mount}
\\
3D printed casing and sensor mounted around the carotid artery of a pig.}
\label{fig:fig4}
\end{figure}

The sensors were then trialled in an acute setting by wrapping the blood pressure sensor around the common carotid artery of anaesthetized female pigs (Fig.\ref{fig:fig4}).  Time series blood pressure measurements were recorded by Powerlab and Labchart (ADInstruments, Oxford) from an invasive/external blood pressure sensor together with a surface electrocardiogram as reference measurements.

\begin{figure*}[ht]
\includegraphics[width=\linewidth]{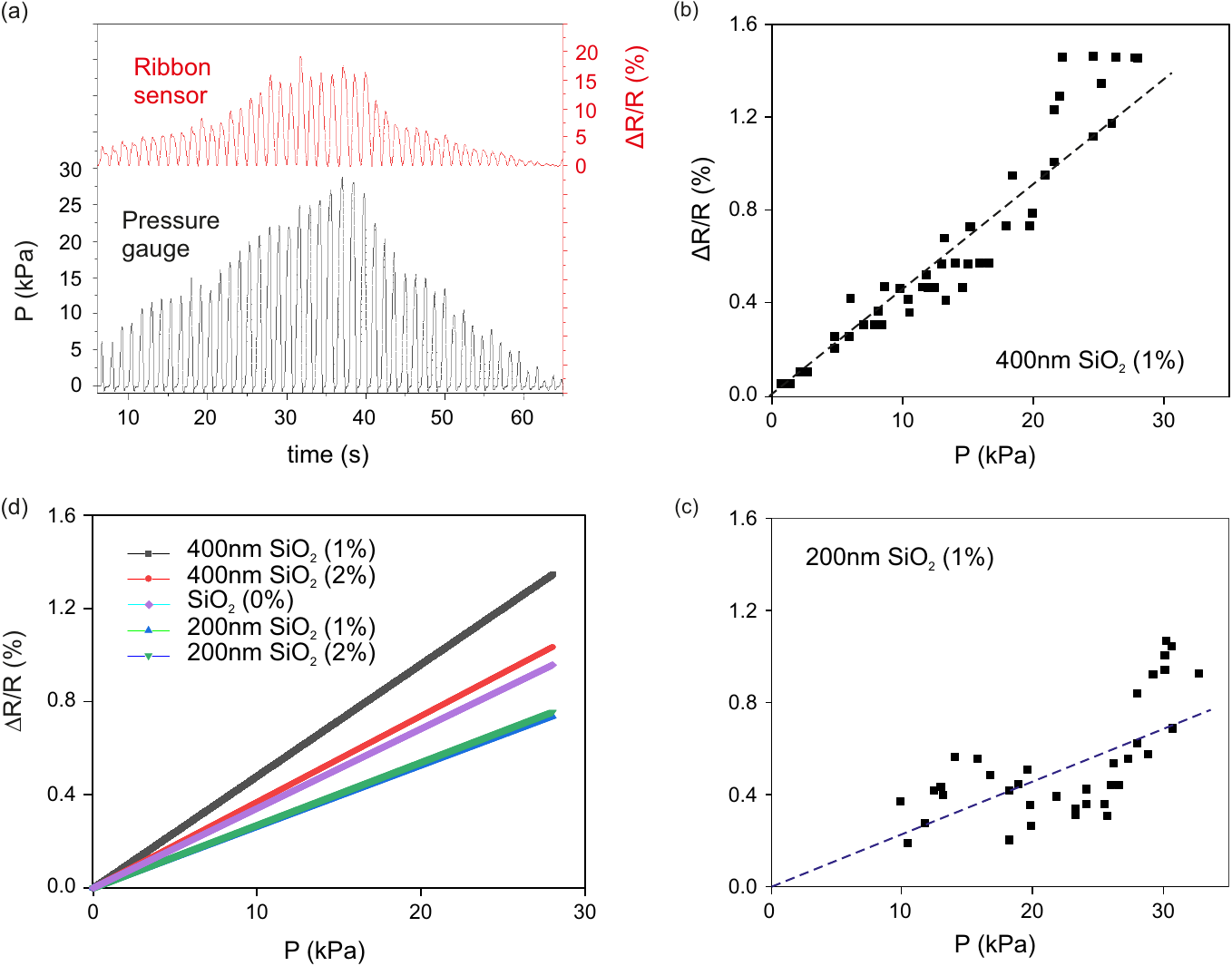}\\
\caption{\textbf{Piezoresistance of bifiller composite nanoribbons}
\\
(a) Piezoresistance measured across the HOPG (29.5\%), 400nm SiO$_2$ (1\%) ribbon (red trace) in response to short pressure pulses obtained by repeatedly compressing the bladder to simulate blood pressure oscillations.  The pezoresistance is calibrated with a 2000hPa pressure gauge (black trace).  (b) Amplitude of piezoresistance peaks plotted as a function of pressure for the HOPG (29.5\%), 400nm SiO$_2$ (1\%) ribbon; (c) Same for the HOPG (29.5\%), 200nm SiO$_2$ (1\%) ribbon.  (d) Linear interpolation of the piezoresistance of the 5 ribbon sensors mixing different concentrations and sizes of SiO$_2$ nanoparticles in PDMS with 29.5\% HOPG.  Sampling frequency: 5Hz.}
\label{fig:fig5}
\end{figure*}

\section{Results}

\subsection{Piezoresistance, response time, incomplete recovery}

A typical sequence of 50 piezoresistance ($\Delta R/R$)and pressure ($P$) oscillations is shown in Fig.\ref{fig:fig5}a.  The average period was 1.5s approximating the R-R cardiac interval at rest.  At 1\% SiO$_2$ (400nm) the piezoresistance peaks lag the pressure peaks by $\tau \approx 0.2$s  This delay relates to the viscoelastic stress relaxation time of the composite.  Fig.\ref{fig:fig5}a (red trace) shows the good reproducibility of the resistance which returns to within 1.2\% of the initial baseline at the end of the cycle.  Fig.\ref{fig:fig5}a shows that the peak resistance changes are broadly proportional to the pressure amplitudes at the peak (black trace).  For each peak the resistance was plotted as a function of pressure in Fig.\ref{fig:fig5}b and Fig.\ref{fig:fig5}c.  The piezoresistance is broadly linear with pressure with some dispersion associated with the uncertainty on the measurement of the peak amplitudes.  Increasing the concentration of 400nm SiO$_2$ nanospheres from 1\% to 2\% leads to a 20\% lower piezoresistance.  Substituting the 400nm SiO$_2$ nanospheres (Fig.\ref{fig:fig5}b) with the 200nm ones (Fig.\ref{fig:fig5}c) decreases the piezoresistance by 50\%.  A summary of the linearly extrapolated piezoresistances for the 5 types of composites is plotted in Fig.\ref{fig:fig5}d.  The largest piezoresistance for the 1\% 400nm SiO$_2$, 29.5\% 450nm HOPG composite.

The respective piezoresistance values for each composite are listed in Table~\ref{tab:tab2} together with the response times, $\tau$, and the incomplete recovery ratio, $\epsilon$.  A most significant result is that increasing the volume fraction of SiO$_2$ not only increases the piezoresistance but dramatically improves the reproducibility of electrical properties from $\epsilon=38\%$ without SiO$_2$, through 22.22\% at 1\% 200nm SiO$_2$, to 1.197\% at 2\% SiO$_2$.  The only tradeoff for inserting SiO$_2$ nanoparticles is a marginally longer recovery time, rising from 0.2s without SiO$_2$ to 0.3s at 2\% 400nm SiO$_2$.

\begin{table*}[ht]
\begin{center}
\begin{tabular}{|c|c|c|c|c|c|}
\hline \hline
& \multicolumn{5}{c|}{Electric properties} \\
& \multicolumn{5}{c|}{PDMS + 29.5\% HOPG 450nm +}  \\
& \multicolumn{2}{c|}{SiO$_{2}$ 400nm} & \multicolumn{2}{c|}{SiO$_{2}$ 200nm} & - \\
& \;\; 1\% \;\; & \;\; 2\% \;\; & \;\; 1\% \;\; & \;\; 2\% \;\; & \; \; - \; \; \\
\hline
$\Delta R/R$ ($\times 10^{-5}$ \%/Pa) & $4.81 \pm 1.13$ & $4.33 \pm 0.253$ & $2.255 \pm 0.139$ & $2.254 \pm 0.185$ & $3.988 \pm 0.174$ \\
$\tau$ (s) & 0.380 & 0.300 & 0.314 & 0282 & 0.205 \\
$\epsilon$ (\%) & 1.524 & 1.197 & 1.770 & 22.22 & 38.243 \\
\hline \hline
\end{tabular}
\end{center}
\caption{\textbf{Piezoresistance, response time, and incomplete recovery of the 5 composites}}
\label{tab:tab2}
\end{table*}

\subsection{Animal trial results}

\begin{figure*}[ht]
\includegraphics[width=\linewidth]{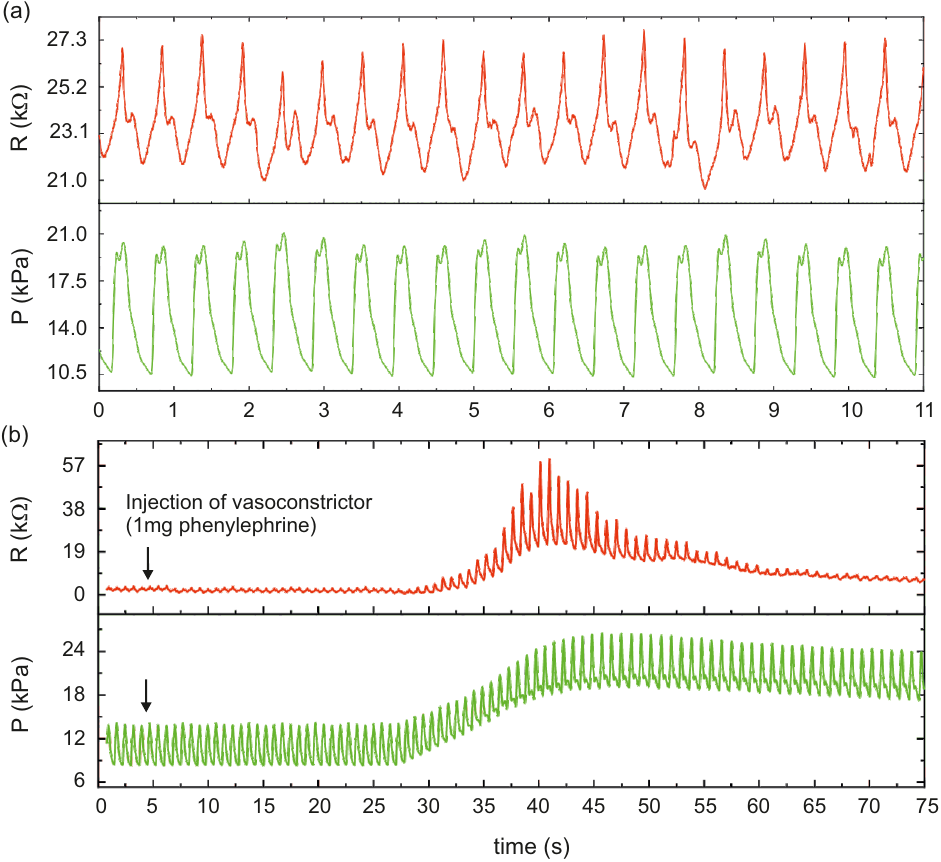}\\
\caption{\textbf{In-vivo recordings}
\\
(a) Time series recordings of spirometry (top trace), externally measured blood pressure (middle trace), and piezoresistance of the ribbon sensor on the carotid artery (bottom trace).  The blood pressure and piezoresistance show the two-peak structure of the systolic and diastolic phases separated by the dicrotic notch.  (b) Blood pressure recording taken after transcutaneous injection of a vasoconstrictor.  This increases the baseline blood pressure as detected by both the external sensor and the piezoresistive ribbon (bottom trace) after a 25s lag time.}
\label{fig:fig6}
\end{figure*}

Fig.\ref{fig:fig6}a shows blood pressure measurements recorded on the carotid artery with the piezoresistive ribbon (Fig.\ref{fig:fig4}).  The ribbon sensor (red trace) detects a double peak in each heart beat which are known as the ejection peak at the end of systole and the isovolumic relaxation peak at the onset of diastole.  These peaks are likewise detected by the external blood pressure sensor (green trace).   However the piezoresistance peak have a higher prominence and are better resolved.  In particular the first larger peak concurs with the larger spike in blood pressure expected during the ejection phase.   Note the 0.2s response delay in the piezoresistance signal relative to the actual blood pressure.  This delay remains small compared to the 1.1s R-R interval.  We also note that the ribbon detects a cyclical modulation of the blood pressure amplitude with a 3-4s period.  These oscillations are in phase with the spirometry signal which we used to monitor respiration.  Hence the ribbon sensor appears to detect a modulation of blood pressure by respiration which is known to occur~\cite{Menuet2020} but is not as obvious in the external blood pressure sensor.

In Fig.\ref{fig:fig6}b, we investigated the response of the sensor to a transient increase in blood pressure by injecting a vasoconstrictor (1mg phenylephrine).  Following a 25s second delay, the amplitude and baseline of the resistively detected pressure increases, and the double peak separation vanishes as validated by the external sensor.  The sensor therefore showed a good correlation of the piezoresistance with the short and longer term changes in blood pressure.

\section{Discussion}

We have shown that minute additions of SiO$_2$ nanoparticles to PDMS nanocomposites significantly improves the reversibility of resistance recovery, increases the piezoresistance hence the detection sensitivity, while modestly increasing the response delay.  We now interpret these results and comment on the temperature dependence of the bi-filler composite.

Starting with the reversibility of the resistance recovery, previous works have considered hydrostatic pressure applied to hard core/soft shell nanoparticles~\cite{Grest1986}.  Our work has studied the effect of uniaxial pressure on a conductive elastomer.  This scenario is more frequent in real-world applications as the conductive elastomer has electrical contacts and stands on a substrate that transmits uniaxial pressure.  Increases in blood vessel circumference applies radial pressure while stretching the ribbon in transverse directions at constant volume.  The two transverse directions are nominally equivalent with a Poisson ratio of 0.5 in each transverse direction.  In practice, this symmetry is easily broken by incomplete stress relaxation ($\epsilon>0$) in non-ideal elastomers~\cite{Hunter1976,Taylor2017}.  This irreversibly alters the topology of the percolation network.  As we have shown, each HOPG nanoparticle constitutes one percolation node where $\geq 3.2$ percolation paths meet in average.  The most tortuous percolation paths are the most susceptible to irreversibility.  Inserting SiO$_2$ nanospheres reduces the number of nearest neighbours which primarily switches off the longest percolation paths.  We interpret the improvement in resistance recovery in all four bi-filler composites to the shutdown of transverse electrical pathways by SiO$_2$ nanospheres (Table~\ref{tab:tab2}).

The increase in piezoresistance of bi-filler composites pushes the effective HOPG filling fraction $\nu$ towards the percolation threshold $\nu_c$.  As a result, the resistance becomes increasingly sensitive to the applied strain and is largest at the 1\% filling fraction of 400nm SiO$_2$ nanospheres (Table~\ref{tab:tab2}).

The 0.2s response time of the HOPG composite is consistent with the viscoelastic relaxation time $\tau_R=\eta/E$ and the creep time $\tau_c=2\tau_R$, where $\eta$ is the viscosity and $E$ the Young's modulus.  The tabulated values for cured RTV137 silicone give $\eta=0.2\pm0.03$MPa.s and $E=2\pm0.2$MPa which yields $\tau_R=0.1$s and $\tau_c=0.2$s in agreement with experiment.  Silica concentrations 5-20 times larger than those we have used here have been shown to increase the Young's modulus~\cite{Ibrahim2012}, hence a decrease in response time might eventually be expected at higher SiO$_2$ filling fractions.

On the bench, the resistance of bi-filler composites varies significantly with temperature as temperature activates percolation hopping.  However, this dependence did not affect our trials in any way as the ribbon was kept at a constant body temperature.

\section{Conclusions}

This study demonstrates the advantages of bi-filler composites in improving the reproducibility and detection sensitivity of \textit{in-vivo} pressure sensors.  We have qualitatively justified the greatly improved reproducibility of our sensors by arguing that insulating nanospheres decouple electrical properties from irreversible changes in the topology of the percolation network associated with incomplete viscoelastic relaxation.  Monte-Carlo simulations may further validate and refine this picture.  Bi-filler composites considerably simplify the manufacture of piezoresistive sensors which previously required complex micro-texturing of fibres and platelets in the PDMS matrix.  The low manufacturing cost of bi-filler composites is expected to increase their availability for \textit{in-vivo} pressure sensing.
\\
\\
\textbf{Author contributions} \\
CB synthesized the sensors and performed the piezoresistive measurements on the benchtop, MR performed sensor trials in pigs, SA designed the 3D printed clamp, MG organised animal trials including ethics compliance, AN conceived the work, secured funding and wrote the manuscript.  All authors contributed to the final version.
\\
\\
\noindent \textbf{Acknowledgments}  \\
This work was supported by the European Union's Horizon 2020 Future Emerging Technologies Programme (Grant No. 732170).
\\
\\
\noindent \textbf{Ethics}  \\
The work has been conducted under ethical approval 337/115-97/98 from the Medical University of Vienna and the Bunderministerium f\"{u}r Bildung, Wissenschaft und Forschung of Austria.
\\
\\
\noindent \textbf{Competing statement}  \\
The authors declare no competing interests.


\bibliography{Sensor_BP_biblio}

\end{document}